# Twisting of light around rotating black holes

Fabrizio Tamburini[1], Bo Thidé[2], Gabriel Molina-Terriza[3] and Gabriele Anzolin[4]

[1]*Department of Astronomy, University of Padova, vicolo dell'Osservatorio 3, I-35122 Padova, Italy, EU*
[2]*Swedish Institute of Space Physics, Box 537, Ångström Laboratory, SE-75121 Uppsala, Sweden, EU*
[3]*QsciTech and Department of Physics and Astronomy, Macquarie University, 2109 NSW Australia*
[4]*ICFO, Parc Mediterrani de la Tecnologia, Av. del Canal Olímpic s/n, ES-08860 Castelldefels (Barcelona), Spain, EU.*

**Kerr black holes are among the most intriguing predictions of Einstein's general relativity theory[1,2]. These rotating massive astrophysical objects drag and intermix their surrounding space and time, deflecting and phase-modifying light emitted nearby them. We have found that this leads to a new relativistic effect that imposes orbital angular momentum onto such light. Numerical experiments, based on the integration of the null geodesic equations of light from orbiting point-like sources in the Kerr black hole equatorial plane to an asymptotic observer[3], indeed identify the phase change and wave-front warping and predict the associated light-beam orbital angular momentum spectra[4]. Setting up the best existing telescopes properly, it should be possible to detect and measure this twisted light, thus allowing a direct observational demonstration of the existence of rotating black holes. Since non-rotating objects are more an exception than a rule in the Universe, our findings are of fundamental importance.**

In curved space-time geometries, the direction of a vector is generally not preserved when parallel-transported from one event to another, and light beams are deflected because of gravitational lensing. If the source of the gravitational field also rotates, it drags space-time with it, causing linearly polarized electromagnetic radiation to undergo polarization rotation similar to the Faraday rotation of light in a magnetized medium. This is known as the gravitational Faraday effect[5]. Because of the rotation of the central mass, each photon of a light beam propagating along a null geodesic will experience a well-defined phase variation. As a result, the beam will be transformed into a superposition of photon eigenstates, each with a well-defined value $s\hbar$ of spin angular momentum and $l\hbar$ of orbital angular momentum[6]. Patterns drawn by such beams experience an anamorphosis[7] with polarization rotation due to the gravitational Faraday effect, accompanied by image deformation and rotation due to the gravitational Berry phase effect[8,9]. Hence, light propagating nearby rotating black holes experiences behaviour analogous to that in an inhomogeneous, anisotropic medium in which spin-to-orbital angular momentum conversion occurs[10].

While the linear momentum of light is connected with radiation pressure and force action, the total angular momentum $J = S + L$ is connected with torque action. The spin-like form $S$, also known as spin angular momentum (SAM), is associated with photon helicity and hence with the polarization of the light. The second form, $L$, is associated with the orbital phase profile of the beam, measured in the direction orthogonal to the propagation axis, and is also known as orbital angular momentum (OAM). This physical observable, which is present in natural light, finds practical applications in nanotechnology[11], communication technology[12] and many other fields. In observational astronomy, OAM of light[13-15] can improve the resolving power of diffraction-limited optical instruments by up to one order of magnitude for non-coherent light[16] and facilitate the detection of extrasolar planets[17-18].

As is well known from quantum electrodynamics[19], each individual photon carries an amount of SAM, quantized as $S = \sigma\hbar$, $\sigma = \pm 1$, and can additionally carry an amount of OAM, quantized as $L = l\hbar$, $l = 0, \pm 1, \pm 2, ..., \pm N$. The OAM of photons has been confirmed experimentally[20,21] and discussed theoretically[22]. Generally, it is not always possible to split the total angular momentum $J$ of a photon into two distinct gauge-invariant observables $S$ and $L$. However, when a paraxial beam of light propagates in vacuum along the $z$ axis, one can project $S$ and $L$ onto this propagation axis and obtain two distinct and commuting operators, $S_z$ and $L_z$, such that

$$\hat{S}_z = \hbar \sum_{\sigma,l,p} \sigma \int_0^\infty dk_0 \hat{a}^\dagger_{\sigma,l,p}(k_0) \hat{a}_{\sigma,l,p}(k_0)$$

$$\hat{L}_z = \hbar \sum_{\sigma,l,p} l \int_0^\infty dk_0 \hat{a}^\dagger_{\sigma,l,p}(k_0) \hat{a}_{\sigma,l,p}(k_0)$$

(eq. 1)

where $\hat{a}_{\sigma,l,p}(k_0)$ and $\hat{a}^\dagger_{\sigma,l,p}(k_0)$ are the creation/annihilation operators of the EM field, expressed in spin ($\sigma$), orbital ($l$) and radial ($p$) angular momentum states of a helical beam. Here, each photon propagation state can be approximated by a Laguerre-Gaussian mode with indices *l* and *p*, i.e. an EM field with amplitude

$$U_{l,p}^{L-G}(r,\vartheta) \propto \left(\frac{r\sqrt{2}}{w}\right)^{|l|} L_p^l\left(-\frac{r^2}{w^2}\right) \exp\left(-\frac{r^2}{w^2}\right) \exp(-il\vartheta) \quad \text{(eq. 2)}$$

where the azimuthal index *l* describes the number of twists of the helical wave front, *p* the number of non-coaxial modes, *w* a scale parameter, $L_p^l(x)$ the associated Laguerre polynomial, and $\vartheta$ the phase.

In geometric units ($G = c = 1$), Kerr space-time, expressed in the Boyer-Lindquist coordinates ($t, r, \theta, \phi$), is described by the line element[1]

$$ds^2 = \frac{\rho^2}{\Delta}dr^2 + \rho^2 d\theta^2 + \frac{\sin^2\theta}{\rho^2}\left[a\,dt - (r^2+a^2)d\phi\right]^2 - \frac{\Delta}{\rho^2}(dt - a\sin^2\theta\,d\phi)^2 \quad \text{(eq. 3)}$$

where the quantities $\rho^2 = r^2 + a^2\cos^2\theta$ and $\Delta = r^2 - 2Mr + a^2$ depend on the mass *M* and on the angular momentum per unit mass *a* ($\leq 1$), i.e. the rotation parameter of the Kerr black hole (KBH). In Kerr metric, orbits are not planar and the only way to calculate the null geodesic equations is to use either the Walker-Penrose conserved quantities or two constants of motion, $\lambda$, and $Q$, that are related to the *z* component of the KBH angular momentum and to the square of the total angular momentum, respectively. Each null geodesic is identified by its impact parameters $\alpha$ and $\beta$ that describe the direction with respect of the image plane of an asymptotic observer located at a latitude $\theta_{obs}$ with respect to the KBH,

$$(\alpha,\beta) = \left(-\frac{\lambda}{\sin\theta_{obs}}, \sqrt{Q + a^2\cos^2\theta_{obs} - \lambda^2\cot^2\theta_{obs}}\right) \quad \text{(eq. 4)}$$

Accretion is thought to occur mainly in the equatorial plane of the KBH[2]. To image, at infinity, the shape of equatorial orbits around a KBH, assumed to model a thin accretion disk, and to calculate the phase acquired by light emitted from that accreting matter, we solve numerically the null geodesic equations in strong gravity conditions by using the software described in Ref. 3 and in the Supplementary Information (SI). The phase variation map of photons emitted by source elements in a region of 100×100 Schwarzschild radii is calculated by using conventional projection techniques. From this map it is straightforward to estimate the OAM spectrum emitted by the radiating matter in that region of the sky. The phase and OAM acquired are independent of both frequency and intensity and the OAM spectrum will be given by the convolution of the acquired OAM and the emission law of the AD. Because of space-time dragging (Lense-Thirring effect), all the sources around the black hole are forced to rotate and the phase of light will change[8,9]. In geometric optics the analogy between the propagation of light in inhomogeneous media and in curved space-times is well established[23]. In this picture, the polarisation and image rotation are attributed to SAM and OAM, respectively. The work Letter extends this to Kerr space-times.

Specifically, we have considered the supermassive Galactic Centre KBH Sgr A*, whose rotation is still a matter of debate: $0.5 \leq a < 0.9939^{+0.0026}_{-0.0074}$ (see Refs. 24 and 25). Figure 1 displays the phase map of light emitted in the equatorial plane around the KBH, projected onto the observer's sky plane of view, for the quasi-extremal case $a = 0.99$ when the inclination is $i = 45°$ with respect to the observer. Owing to the asymmetric gravitational lensing distortion and the black hole rotation, this light has quite a wide and structured OAM spectrum (upper right-hand inset). Figure 2 shows, for the representative cases $a = 0.99$ (top) and $a = 0.5$ (bottom), the effect of the Sgr A* rotation on the photon phase (left), normalised to a quasi-static ($a=0.01$) KBH, inclined by the same angle, and the corresponding OAM spectra (right). In both cases the morphing effects due to a particular inclination of the Sgr A* equatorial plane have been removed in order to exhibit the purely Kerr metric effects. Our simulations show that the main contribution to the phase difference comes from the inner stable orbits that approach the Sgr A* event horizon and that slowly rotating black holes give rise to uniformly decaying power distributions of OAM modes around a zero OAM value.

In order to detect rotating black holes with the technique described here, it is sufficient to use the best available telescopes, provided that they are equipped with proper OAM diagnostic instrumentation (e.g. holographic detectors). As OAM of light is merely related to its spatial structure, only spatial coherence of the source is required[18,26]. The long distances from the KBH to the observer ensure this spatial coherence, even if the electromagnetic fields generated at different points in the AD are mutually incoherent (see SM). Additional information about the KBH rotation can be obtained by analysing the OAM spectra in different frequency bands of the electromagnetic radiation as the new gravitational effect described in this Letter is wavelength independent. Since we are only interested in the gravitational effect, we have not taken into account the possibility that interactions with plasma turbulence may induce variations in the OAM spectrum of the light at certain frequencies[27]. However, these variations are likely to be much smaller than the gravitational effect. We also anticipate that the analysis of several different frequency bands will provide useful information about the turbulence of the disk corona and of the traversed interstellar medium, as well as the pure gravitational effect.

The observables associated with a Kerr metric considered here are invariant with respect to the mass of the KBH. Consequently, the photon phase $\vartheta$ and the concomitant OAM (spiral) spectrum[4] will depend only on the rotation parameter $a$ and on the inclination $i$ of the equatorial plane relative to the observer.

Contributions from the secondary ghost images[3,8,9,28] generated, by those light beams that wind their paths $n$ times around a KBH, are expected to widen the OAM spectra. However, as $n$ increases, the secondary ghost images become fainter and more difficult to observe[7]. Moreover, higher-order correction terms would lead to photon spin precession, a negligible second-order effect in the Lagrangian[8]. An investigation into these effects goes beyond the scope of this Letter. High OAM values may also occur in particular situations of radial polar accretion[29] or from the gravitational lensing of coherent astrophysical sources. The OAM mechanism described here should be valid also for neutrino fields[30]. Our results can be extended to more general situations such as in the latest stages of BH-BH collisions or in generalizations of Kerr metrics.

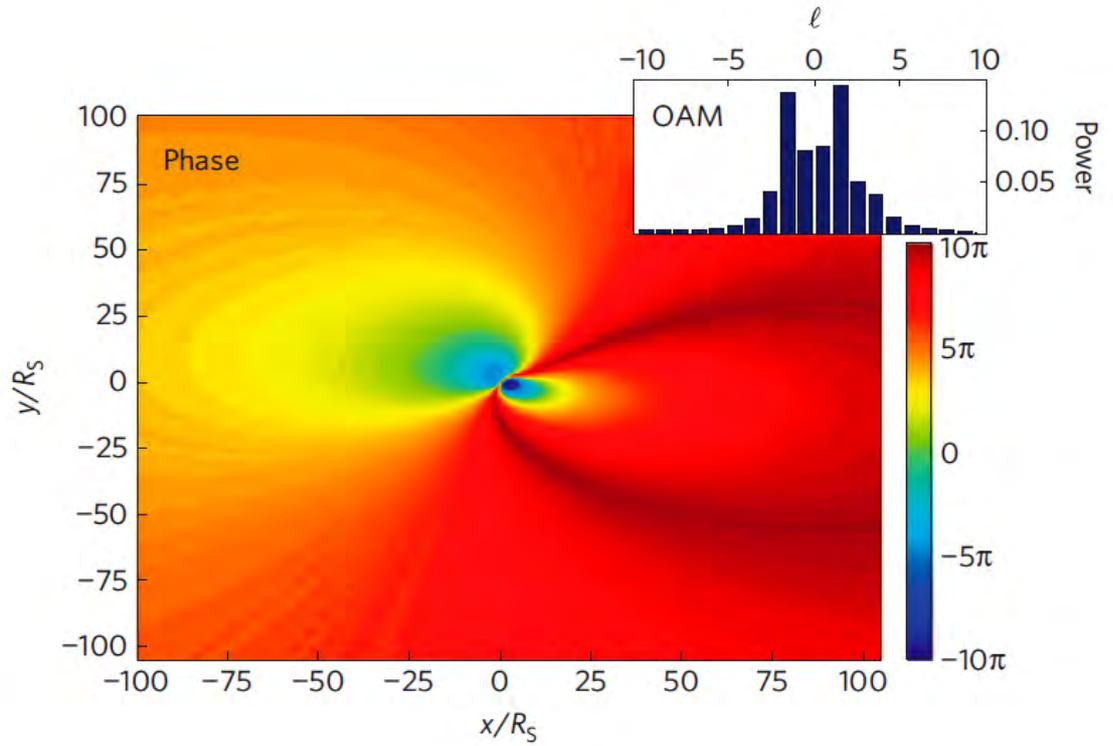

**Fig. 1**

Total phase variation of light generated in a region of 100 times 100 Schwarzschild radii ($R_s$) in the equatorial plane of a quasi-extremal rotating black hole ($a = 0.99$) as seen by an asymptotic observer located at infinity. This portion of the sky shows what would be observed with a telescope if the black hole rotation axis is inclined an angle $i = 45°$ relative to the observer. The total phase variation includes the *anamorphic* effect due to both the space-time curvature and the inclination of the disk. As shown in the right-hand panel, the corresponding OAM spectral distribution is quite complex, with two strong peaks at $\ell = -2$ and $\ell = 1$, and extends towards higher OAM modes with a rapid fall-off.

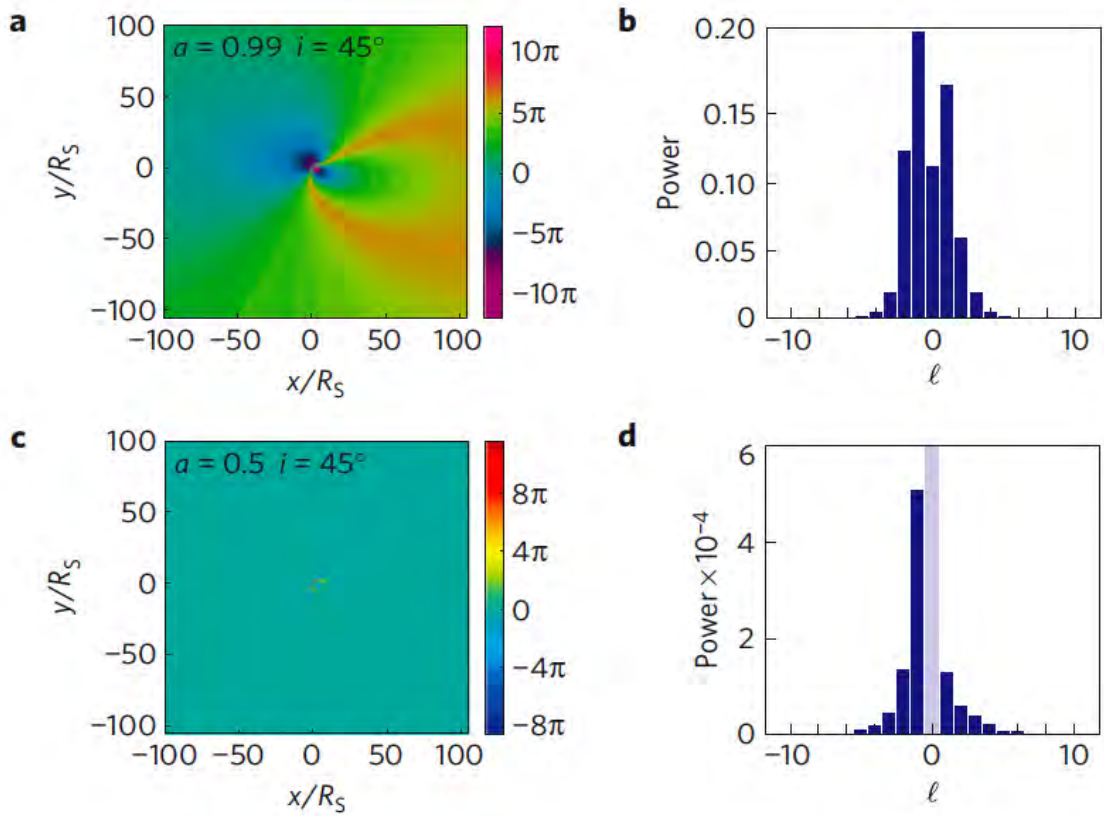

**Fig. 2**

Phase variation of photons as measured by an asymptotic observer. The $(x, y)$ plane represents a 100 by 100 Schwarzschild radii large region of the sky centred on the KBH. The two plots in panels A and C depict the OAM acquired due only to the KBH rotation for $a$ = 0.5 and $a$ = 0.99, respectively, normalised to the field of a quasi-static black hole ($a$ = 0.01). Here we estimate the torsion of the optical path due to the space-time dragging of the KBH. For a KBH with $a$ = 0.5, the only significant contribution is a narrow OAM spectrum that comes from the immediate neighbourhood of the compact object where the relativistic effects are strongest. In contrast, the space-time dragging of the extremal BH ($a$ = 0.99) shows a bimodal OAM power spectrum distribution peaked at $\ell = -1$ and $\ell = 1$ relative to a static BH. The torsion is zero if the black hole is static, in agreement with Ref. 9. The panels B and D display the OAM spectra of the cases A and C, respectively. While the OAM spectrum in panel B has its maximum power in the $\ell = 1$ mode, the most powerful mode in panel D is the $\ell = 0$ mode. For this reason we plot in this case only a magnification of the residual power that carries non-vanishing OAM.

# Appendix

## 1. Radiation from a thin accretion disk around a Kerr black hole observed remotely at asymptotic distances

In our numerical simulations we model the radiation from a thin accretion disk (AD) around a Kerr black hole (KBH), as seen by a remote observer who is located at an asymptotic distance that is much larger than the maximum extent of the AD itself. Specifically, we investigate numerically the twisting of the light caused by the central black hole of our Galaxy, Sgr A*. For a detailed account we refer to Ref. 26.

In the Kerr metric, the KBH is located at the centre of the ($x, y, z$) coordinate system. The $z$ axis coincides with the rotation axis of the KBH and – in the simplest case – the thin unperturbed AD is located in the orthogonal ($x, y$) plane. For the sake of simplicity, more complicated accretion geometries such as tilted disks, thick disks, tori, etc. are not considered.

In the Boyer-Lindquist (B-L) coordinate system used, the analogy with spherical coordinates, of which Kerr geometry is a generalization, is evident. Each of the observers is identified by an event, ($t, r, \theta, \varphi$), in the four-dimensional manifold. The local Galilean reference frames associated with each of the asymptotic observers are Cartesian and rotated with respect to the $z$ axis of the B-L coordinates. Their $z'$ axes coincide with the line of sight from the observer to the black hole. The orthogonal plane ($x', y'$) of the observer is the "sky view" of a region of the sky surrounding the KBH. In more practical terms, this is what one can image with a telescope pointed toward the galactic centre. The geometry is outlined in Figure 3.

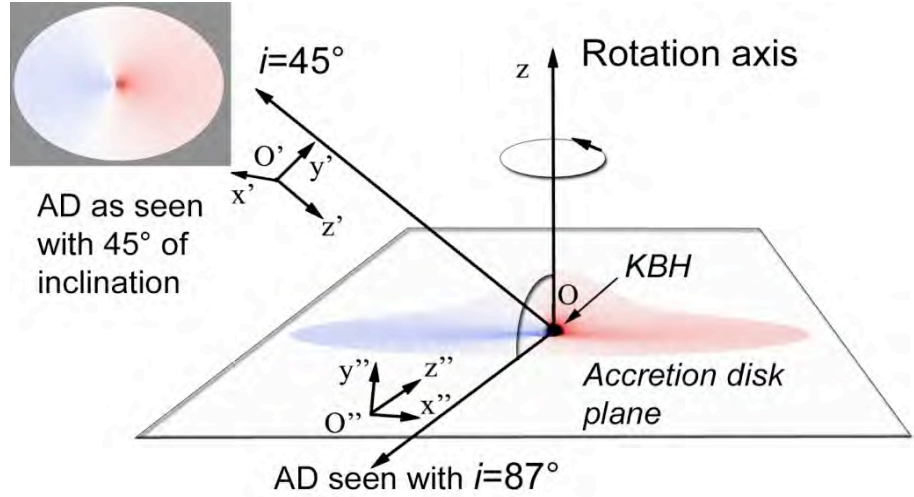

**Figure 3: Thin accretion disk as seen by an observer at infinity at *i*=45° (*x'*, *y'*, *z'*) and at *i* = 87° (*x''*, *y''*, *z''*).**

Because of gravitational lensing, each observer will see a different deformed pattern of the accretion disk. The pattern will mainly depend on the inclination parameter *i* of the observer with respect to the rotation axis *z*, and on the rotation parameter *a* of the KBH. It is equivalent to saying that the KBH is inclined with different angles with respect to the line of sight of a fixed observer. In Figure 1 we also show an example of the deformed pattern of a thin planar AD, when *i* = 45° and when the observer is almost orthogonal to the KBH rotation axis, *i* = 87°, where the deformation of the thin accretion disk due to the gravitational lensing is more evident.

## 2. Numerical integration of the null geodesic equations

In order to solve the null geodesic equations and calculate the associated photon phase in the Kerr geometry for each geodesic that connects a point-like source in the equatorial plane of the KBH with an asymptotic observer, we used a fast and accurate numerical method based on the Jacobian elliptical function approach. The main core of the software, developed by Čadež and Calvani [3], is freely available. This software was written to simulate the spectral line profiles emitted by an AD around a Kerr black hole [3,31]. In our simulations we focused the attention on the simplest case of a planar, thin AD. Other available options that analyze the emission of inclined, twisted, and planar (flat) disks will be the subject of a future investigation. In the model that we consider in the Letter, the distribution of the orbital angular momentum states of light is caused only by the gravitational effects of the KBH and is not influenced by the distortion of the AD.

The physical parameters of the AD in our simulations were the following:

> $R_{out}$ (outer disk radius) = 50 $R_s$, where, in natural units, the Schwarzschild radius $R_s$ = 2$M$, and $M$ is the KBH mass, i.e. $R_{out}$ = 100$M$.
>
> $\Theta$ (disk inclination angle with respect to line of sight) = 45°.
>
> $a \times 100$ (angular momentum of BH) = 1$M$ (i.e. $a$ = 0.01$M$); $a$ = 0.5 or $a$ = 0.99.

The emissivity law can be approximated by a power law, $\varepsilon(r) \propto r^{-q}$, where $r$ is the orbital fiducial radius of the emitting particle forming the AD and $q$ the exponent of the power law. Alternatively, the illumination of the disk can be light emitted by one or two point-like sources on the axis of the disk at a given height $\pm h$ on the AD plane. More precisely the emissivity law $\varepsilon_{TOT}(r,\mu)$, where $\kappa$ is the cosine of the angle of the outgoing light ray with respect to the disk normal vector, can be factorized into two terms $\varepsilon(r)$ and $F(\mu)$, according to the general emissivity formula $\varepsilon_{TOT}(r,\mu) = \varepsilon(r)F(\mu)$.

The radial emissivity factor $\varepsilon(r)$ is expressed by

$$\varepsilon(r) = K \left( \frac{1+n_2}{1+n_1} \right) \left( \frac{1+n_1\mu}{1+n_2\mu} \right) \left( \frac{r}{r_{min}} \right)^{-q}.$$

By varying the two parameters $n_1$ and $n_2$ and the power-law index $q$, one can describe a wide variety of situations related to the AD emission. A pure power law emissivity without angular dependence is given by $F(\mu) = 1$, the limb darkening law is obtained by setting $F(\mu) = 1 + 2.06\mu$, and the limiting case of optically thin material corresponds to $F(\mu) \propto 1/\mu$, in agreement with Ref. 26. A different intensity emissivity law of a symmetric thin equatorial disk does not change the profile of the OAM spectrum, since the OAM states are independent of the intensity distribution of the source.

In our simulations, we have also tested the dependence of the OAM spectrum obtained by the emission law of the AD by varying the parameters that determine the intensity line profile emission of the accretion disk. As expected from the achromaticity of this phenomenon, no particular variations in the phase distribution were observed and, consequently, no appreciable changes in the OAM spectra were found.

## 3. The Orbital Angular Momentum Spectrum

As already discussed, the orbital angular momentum (OAM) states of light are particular photon states that are often represented in terms of corkscrew patterns describing equal-phase regions about the beam axis [6].

For an asymptotic observer to be able to detect and analyze OAM of light from a star or, equivalently, produce optical vortices with starlight, the following conditions must be fulfilled:

The detrimental effects of atmospheric turbulence must be eliminated by using either a diffraction-limited telescope, such as Hubble space telescope, or a telescope with an efficient adaptive optics system.

Spatial coherence (see Ref. 18). This is ensured by the geometry of the problem where the light arriving from a star is spatially coherent, giving rise to diffraction at the telescope aperture. Spatial coherence of light, from an astronomer's point of view, means that Airy diffraction patterns are obtained when a star is imaged with a diffraction-limited telescope.

It should be noted that optical vortices can be present in incoherent and polychromatic light, such as light from stars that emit like black bodies. As shown in the literature cited in Ref. 18, it is not mandatory that the source is temporal coherent or photon number coherent, in order to produce OAM. In fact, it is well known that photons received from astronomical sources follow a Poissonian distribution in time. This process was mimicked in the laboratory with a white light lamp in order to produce OVs to calibrate our instrument used in Ref. 18.

Even if the radiation sources in the accretion disk around black holes are not coherent but subject to thermal or particular emission laws, this does not alter the distribution of OAM states acquired from the black hole. The OAM states depend only on the trajectory of the light beam and not on the wavelength. Our results show that the effects on the phase of photons from the Kerr black hole geometry are independent of the wavelength. The only (inessential) constraint on the wavelength is to be smaller than the size of the black hole area. This means that practically all wavelengths are affected by the gravitational field in the same way as by a perfect achromatic phase mask in a laboratory.

The angular momentum of light collected by the telescope may contain not only a spin contribution, associated with polarization that have suffered a rotation due to the gravitational Faraday effect, but may also contain an orbital contribution associated with the spatial profile of the light beam amplitude and the phase front as in Laguerre-Gaussian beams. In our case we analyzed a thin, planar, uniformly illuminated equatorial AD with a symmetric shape around the centre of the KBH. This oversimplified scenario illustrates the effect of the gravitational field on electromagnetic waves and photons in the simplest way.

Regarding the angular momentum of light, any light beam in the paraxial approximation can be characterized by its spin and angular momentum. The first of these (SAM) can, for instance, be measured with a polarimeter that characterizes the polarization of the source in terms of the Stokes parameters. Independent of the SAM measurement, one can analyze the phase distribution of the light collected from the portion of the sky, centred on the black hole focused by the telescope, to obtain the OAM spectrum, also known as the spiral spectrum.

OAM states are associated only with the topological properties of the light wave front (and are therefore sensitive to phase gradients and discontinuities) but do not depend on the intensity emissivity law of the source. The OAM spectrum will then depend on the OAM acquired and its intensity distribution, and the two effects can, in principle, be studied separately by solving the Einstein equations of light and those of the accretion disk model.

To generate the total spiral spectrum of a realistic AD, one has to take into account also the radial emission intensity law and convolve the basic spiral spectrum with the intensity emissivity. This gives the power law intensity emission of each spectral line, independently of the amount of phase change experienced by each photon.

In Figure 4 we show the results of a simulation of an OAM spectrum generated by an optically thin disk with $F(\mu) \propto 1/\mu$ and that with a limb-darkening profile emission represented by $\varepsilon(r,\mu) = 0.03(1 - 2.06\mu)/r$.

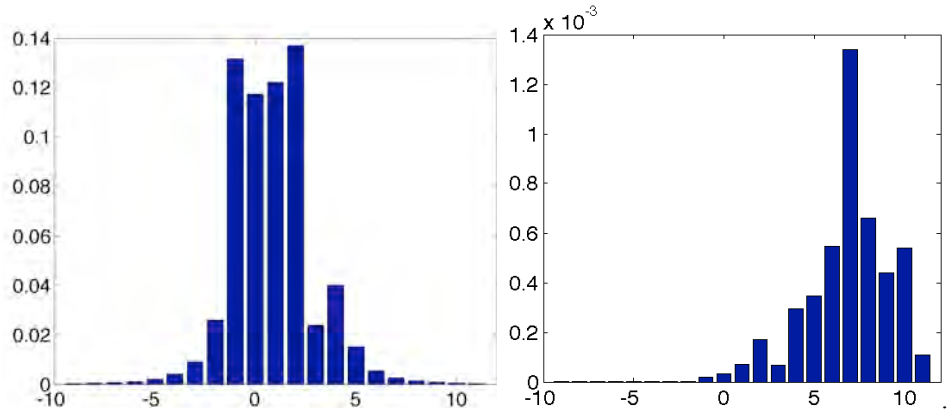

**Figure 4: Spiral spectrum of an optically thin accretion disk and of an accretion disk with limb darkening emission.**

A comparison with the simplest case visualized in Figure 1 suggests that to measure the spin of a KBH, one should take in consideration many different aspects of the accretion phenomena of the object under examination and reduce the spiral spectrum obtained experimentally to the case of a planar uniform AD. One should then consider:

    1. The presence of magnetic fields that can drive the accreting material away from the relativistic geodesics calculated by only taking into account the presence of the gravitational field.

    2. The presence of disk tilting or parametric instabilities in the disk, such as quasi-periodic oscillations, trailing waves, hot spots, common envelope in a binary system, wind accretion and disk instabilities in general.

    3. The emission law of the AD.

    4. The actual geometry of the accreting material. A single star falling into a galactic KBH will be characterized by signatures that are slightly different from those expected from an AD.

    5. The generalization of the Kerr geometry in neutron star –black hole collisions or in lensing phenomena and emissions from BH-BH collisions.

More information about the expected signal-to-noise (S/N) ratio for new and upcoming telescopes can be found in the cited literature. As can be easily concluded from the OAM spectra depicted in Figures 1 and 2 in the Letter, one must collect a certain number of photons. The amount depends on the S/N ratio of the astronomical observations [26] but also on the finiteness of the histogram. The faster the KBH is rotating, the fewer photons are needed to characterize this rotation. Slowly-rotating BHs may need on the order of $10^4 - 10^5$ events to reveal rotation. This can be achieved by integrating over time, if the process is stationary, or by using large aperture telescopes (or both).

OAM spectra can provide useful information in more general cases than the simplest Kerr solution. The latest stages of the collision of compact objects such as neutron star and/or black hole collisions induce similar effects in the space-time metrics that affect the photon phase. Of course, a more detailed numerical analysis is required in order to estimate the phase evolution of photons and the corresponding OAM spectra, but this will be the subject of a future investigation.

### 4 Additional information

In each of our simulations we calculated the phase distribution of the light collected by an asymptotic observer from an area of size 100×100 Schwarzschild radii centred on the KBH. Each of the plots in Figure 1 and Figure 2 are made by mesh plots of approximately 710,000 points for which we calculated the phase and the orbital angular momentum.

For the sake of completeness, we display in the two panels of Figure 5 some additional information that can be obtained with the numerical routine of Ref. 3. The numerical routine calculates also the projection of the unit vectors of the beam propagation (what is usually called a *k*-vector) and the shape of the accretion disk as seen by the asymptotic observer.

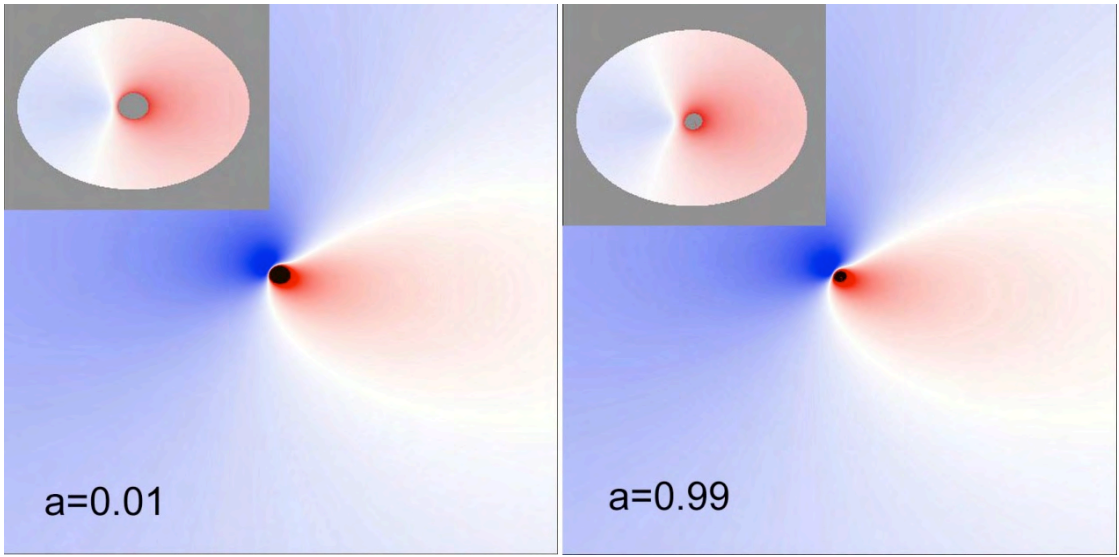

**Figure 5: The two panels visualize, for a static black hole ($a \sim 0$) and for an extremal Kerr black hole ($a \sim 1$), the *k* vector for a thin equatorial accretion disk inclined by $i = 45°$ with respect to the observer.**

The main body of the left-hand panel in Figure 5 displays a color mesh-plot describing the projection of the unit vectors of the beams propagating from the neighborhood of the KBH onto the direction to the observer, namely the values of the projection of the wave vector (*k* vector) with respect to the line of sight. The values of the projections, which vary from -1 to 1, are expressed in the mesh plot with a spectrum of colors that span from red to blue in a discrete scale of 256 values, in an 8-bit scale depth.

The inset of each panel shows the pattern of the corresponding accretion disk, as seen by the distant observer, with an external radius $R_{max} = 40M$, to give better evidence to the regions of the inner last stable orbits that approach the KBH event horizon when the rotation parameter approaches unity. The accretion disk image is color coded according to amount of red-shift (the disk is assumed to rotate from left to right) with a spectrum of colours that span from red to blue. In this case the colour plot represents the Doppler effect due to the gravitational field and the rotation of the matter in the accretion disk. More details can be found in Ref. 3.

## **Acknowledgements**

The authors thank Juan P. Torres, M. Calvani, A. Čadež, and Sir Michael Berry for helpful comments and suggestions . F. T. gratefully acknowledges the financial support from the CARIPARO Foundation within the 2006 Program of Excellence and the kind hospitality of Uppsala University/Swedish Institute of Space Physics and ICFO during the writing of the manuscript. B. T. gratefully acknowledges financial support from the Swedish Research Council (VR) and the hospitality of the Nordic Institute for Theoretical Physics (NORDITA), the University of Padua, and the Institute for Quantum Optics and Quantum Information where parts of this work were carried out.

**Author Contributions**  F. T., B. T. and G. M.-T. developed the model. F. T. performed the numerical simulations. G. A. calculated and plotted the OAM spectra. F. T. and B. T. wrote the manuscript. All authors discussed and commented on the manuscript.